\documentclass[12pt]{nature}

\spacing{2}
\spacing{1}
\usepackage{graphicx}	
\usepackage{amsmath}	
\usepackage{amssymb}	
\usepackage{xcolor}
\usepackage[normalem]{ulem}

\newcommand{\blue}{\textcolor{blue} }
\newcommand{\black}{\textcolor{black} }

\renewcommand{\blue}{\textcolor{black} }

\renewcommand{\sout}[1]{\relax}

\newif\ifarxiv
\newif\ifword

\ifword
    \renewcommand{\blue}{\textcolor{black} }
    \raggedright
    \renewcommand{\echelle}{echelle}
    \arxivfalse
\else
    \arxivtrue
\fi

\usepackage{supertabular}

\newcommand{\apj}{Astrophys. J.}

\newcommand{\apjs}{Astrophys. J. Suppl. Ser.}
\newcommand{\aap}{Astron. Astrophys.}
\newcommand{\aapr}{Astron. Astrophys. Rev.}
\newcommand{\aaps}{Astron. Astrophys. Suppl. Ser.}
\newcommand{\araa}{Ann. Rev. Astron. Astrophys.}
\newcommand{\jaap}{J. Astron. Astrophys.}
\newcommand{\mnras}{Mon. Not. R. Astron. Soc.}
\newcommand{\aj}{Astron. J.}
\newcommand{\nat}{Nature}

\newcommand{\pasp}{Pub. Astron. Soc. Pac.}
\newcommand{\pasa}{Pub. Astron. Soc. Aust.}

\newcommand{\procspie}{Proc. S.P.I.E}

\let\citep\cite
\let\citet\cite

\newcommand{\Teff}{\mbox{$T_{\rm eff}$}}
\newcommand{\kepler}{{\em Kepler\/}}
\renewcommand{\kepler}{Kepler}
\newcommand{\gaia}{{\em Gaia\/}}
\renewcommand{\gaia}{Gaia}
\newcommand{\tess}{{\em TESS\/}}
\renewcommand{\tess}{TESS}
\newcommand{\hipparcos}{{\sc Hipparcos\/}}
\newcommand{\echelle}{{\'e}chelle}

\newcommand{\dsct}{\mbox{$\delta$~Scuti}}
\newcommand{\dsctbf}{$\boldsymbol{\delta}$~Scuti}

\newcommand{\bpic}{\mbox{$\beta$~Pictoris}}
\newcommand{\lboo}{\mbox{$\lambda$~Bo{\"o}tis}}
\newcommand{\cd}{\mbox{d$^{-1}$}}
\newcommand{\kms}{\mbox{km\,s$^{-1}$}}
\newcommand{\muhz}{\mbox{$\mu$Hz}}
\newcommand{\Dnu}{\mbox{$\Delta\nu$}}
\newcommand{\vsini}{\mbox{$v\sin i$}}

\newcommand{\Msun}{\mbox{$M_{\odot}$}}
\newcommand{\Rsun}{\mbox{$R_{\odot}$}}

\newcommand{\rhosun}{\mbox{$\rho_{\odot}$}}
\newcommand{\FeH}{\mbox{[Fe/H]}}

\newcommand{\sfTeff}{\mbox{{\it\sffamily T}$_{\sf eff}$}}
\newcommand{\sfvsini}{\mbox{{\it\sffamily v}\,sin\,{\it\sffamily i}}}
\newcommand{\sfLsun}{\mbox{{\it\sffamily L}$_{\odot}$}}
\newcommand{\sfcd}{\mbox{\sf d$^{-1}$}}
\newcommand{\sfkms}{\mbox{\sf km\,s$^{\sf -1}$}}
\newcommand{\emdash}{\multicolumn{1}{c}{---}}

\title{Very regular high-frequency pulsation modes in young\\ intermediate-mass stars}

\newcommand{\SIfA}{\small\textbf{\sffamily{1}}}
\newcommand{\SAC}{\small\textbf{\sffamily{2}}}
\newcommand{\IfA}{\small\textbf{\sffamily{3}}}
\newcommand{\Bham}{\small\textbf{\sffamily{4}}}
\newcommand{\Keele}{\small\textbf{\sffamily{5}}}
\newcommand{\UNSW}{\small\textbf{\sffamily{6}}}
\newcommand{\ANU}{\small\textbf{\sffamily{7}}}
\newcommand{\Caltech}{\small\textbf{\sffamily{8}}}
\newcommand{\EarthSci}{\small\textbf{\sffamily{9}}}
\newcommand{\LCO}{\small\textbf{\sffamily{10}}}
\newcommand{\BU}{\small\textbf{\sffamily{11}}}
\newcommand{\UCLan}{\small\textbf{\sffamily{12}}}
\newcommand{\Beijing}{\small\textbf{\sffamily{13}}}
\newcommand{\UNC}{\small\textbf{\sffamily{14}}}
\newcommand{\LESIA}{\small\textbf{\sffamily{15}}}
\newcommand{\Leuven}{\small\textbf{\sffamily{16}}} 
\newcommand{\DTU}{\small\textbf{\sffamily{18}}}
\newcommand{\MPI}{\small\textbf{\sffamily{17}}}
\newcommand{\Berkeley}{\small\textbf{\sffamily{19}}}
\newcommand{\Ames}{\small\textbf{\sffamily{20}}}
\newcommand{\Vilnius}{\small\textbf{\sffamily{21}}}
\newcommand{\UCSB}{\small\textbf{\sffamily{22}}}
\newcommand{\MIT}{\small\textbf{\sffamily{23}}}
\newcommand{\KavliMIT}{\small\textbf{\sffamily{24}}}

\author{\small\sffamily\textbf{%
Timothy R.~Bedding$^{\SIfA,\SAC}$\thanks{e-mail: tim.bedding@sydney.edu.au},
Simon J.~Murphy$^{\SIfA,\SAC}$,
Daniel R.~Hey$^{\SIfA,\SAC}$,
Daniel~Huber$^{\IfA}$,
Tanda~Li$^{\SIfA,\SAC,\Bham}$,\\
Barry Smalley$^{\Keele}$,
Dennis~Stello$^{\UNSW,\SAC}$,
Timothy~R.~White$^{\SIfA,\SAC,\ANU}$, 
\black{Warrick~H.~Ball$^{\Bham,\SAC}$,}\\
William J.~Chaplin$^{\Bham,\SAC}$,
Isabel~L.~Colman$^{\SIfA,\SAC}$,
Jim~Fuller$^{\Caltech}$,
Eric~Gaidos$^{\EarthSci}$,
Daniel~R.~Harbeck$^{\LCO}$,\\
J.~J.~Hermes$^{\BU}$,
\black{Daniel~L.~Holdsworth$^{\UCLan}$,}
Gang~Li$^{\SIfA,\SAC}$,
Yaguang Li$^{\SIfA,\SAC,\Beijing}$,
Andrew~W.~Mann$^{\UNC}$,\\
Daniel R.~Reese$^{\LESIA}$,
Sanjay~Sekaran$^{\Leuven}$,
Jie~Yu$^{\MPI}$, 
Victoria~Antoci$^{\SAC,\DTU}$,
\black{Christoph Bergmann$^{\UNSW}$,}\\
Timothy M.~Brown$^{\LCO}$, 
Andrew~W.~Howard$^{\Caltech}$, 
\black{Michael J. Ireland$^{\ANU}$,}
Howard~Isaacson$^{\Berkeley}$,\\
\black{Jon M. Jenkins$^{\Ames}$},
\black{Hans Kjeldsen$^{\SAC,\Vilnius}$},
Curtis McCully$^{\LCO}$, 
Markus Rabus$^{\LCO,\UCSB}$,
\black{Adam D. Rains$^{\ANU}$,}\\
\black{George R. Ricker$^{\MIT,\KavliMIT}$,} 
\black{Christopher G. Tinney$^{\UNSW}$} \&
\black{Roland K. Vanderspek$^{\MIT,\KavliMIT}$}}}

\begin{document}

\maketitle
\begin{affiliations}
\small
\medskip
\item Sydney Institute for Astronomy (SIfA), School of Physics, University of Sydney, Camperdown, New South Wales, Australia.
\item Stellar Astrophysics Centre, Department of Physics and Astronomy, Aarhus University, Aarhus, Denmark.
\item Institute for Astronomy, University of Hawai`i, Honolulu, HI, USA.
\item School of Physics and Astronomy, University of Birmingham, Birmingham, UK.
\item Astrophysics Group, Lennard-Jones Laboratories, Keele University, Keele, UK.
\item School of Physics, University of New South Wales, Kensington, New South Wales, Australia.
\item Research School of Astronomy and Astrophysics, Mount Stromlo Observatory, The Australian National University, Canberra, Australian Capital Territory, Australia.
\item TAPIR, California Institute of Technology, Pasadena, CA, USA.
\item Department of Earth Sciences, University of Hawai`i, Honolulu, HI, USA.
\item Las Cumbres Observatory Global Telescope, Goleta, CA, USA.
\item Department of Astronomy, Boston University, Boston, MA, USA.
\item \black{Jeremiah Horrocks Institute, University of Central Lancashire, Preston, UK.}
\item Department of Astronomy, Beijing Normal University, Beijng, China.
\item Department of Physics and Astronomy, University of North Carolina at Chapel Hill, Chapel Hill, NC, USA.
\item LESIA, Observatoire de Paris, Universit{\'e} PSL, CNRS, Sorbonne Universit{\'e}, Universit{\'e} de Paris, Meudon, France.
\item Instituut voor Sterrenkunde (IvS), KU Leuven, Leuven, Belgium.
\item Max-Planck-Institut f{\"u}r Sonnensystemforschung, G{\"o}ttingen, Germany.
\item \blue{DTU Space, National Space Institute, Technical University of Denmark, Kongens Lyngby, Denmark.}
\item Department of Astronomy, University of California at Berkeley, Berkeley, CA, USA.
\item \black{NASA Ames Research Center, Moffett Field, CA, USA.}
\item \blue{Institute of Theoretical Physics and Astronomy, Vilnius University, Vilnius, Lithuania.}
\item Department of Physics, University of California, Santa Barbara, CA, USA.
\item \black{Department of Physics, Massachusetts Institute of Technology, Cambridge, MA, USA.}
\item \blue{Kavli Institute for Astrophysics and Space Research, Massachusetts Institute of Technology, Cambridge, MA, USA.}
\end{affiliations}


\bigskip

\noindent
(accepted for publication in {\it Nature})
\bigskip
\clearpage

\begin{abstract}
Asteroseismology probes the internal structures of stars by using their natural pulsation frequencies\cite{Aerts++2010-book}.  It relies on identifying sequences of pulsation modes that can be compared with theoretical models, which has been done successfully for many classes of pulsators, including low-mass solar-type stars\cite{Garcia+Ballot2019}, red giants\cite{Hekker+ChD2017}, high-mass stars\cite{Aerts2015} and white dwarfs\cite{Corsico++2019}.  However, a large group of pulsating stars of intermediate mass---the so-called \dsctbf\ stars---have rich pulsation spectra for which systematic mode identification has not \blue{hitherto} been possible\cite{Goupil++2005,Handler2009}.  This arises because only a seemingly random subset of possible modes are excited and because rapid rotation tends to spoil the regular patterns\citep{Ouazzani++2015,Reese++2017,Mirouh++2019}.  Here we report the detection of remarkably regular sequences of \black{high-frequency} pulsation modes in 60 intermediate-mass main-sequence stars, which enables definitive mode identification.  The space motions of some of these stars indicate that they are members of known associations of young stars, as confirmed by modelling of their pulsation spectra.
\end{abstract}

\bigskip

The \dsct\ variables are stars of intermediate mass (1.5--2.5 solar masses, \Msun) that pulsate in low-order pressure modes\citep{Goupil++2005,Handler2009}. 
Observations have shown that many \dsct\ stars have regular frequency spacings in their pulsation spectra (see Methods) but a large sample with unambiguous mode identifications is lacking.  Each pulsation mode in a non-rotating star is identified by two integers: the radial order, $n$, and the degree,~$l$.  We expect the strongest observable modes to be of low degree ($l=0$, 1 and 2), because higher degrees have greatly reduced amplitudes due to cancellation in disk-integrated light.  In the so-called asymptotic regime ($n \gg l$), modes with a given degree $l$ are approximately equally spaced in frequency by a separation, \Dnu, that is the inverse of the time taken for sound waves to travel through the star and is approximately proportional to the square root of the mean stellar density\black{\cite{Aerts++2010-book}}.

The patterns are more complex in a rotating star, with the mode frequencies also depending on the azimuthal order,~$m$.  Each nonradial ($l \ge 1$) mode in the pulsation spectrum is split into $2l+1$ components, where $m$ ranges from $-l$ to~$l$.  The relative amplitudes of these components depend on the inclination of the rotation axis to the line of sight.  For example, if a star is seen at low inclination (close to pole-on) then the axisymmetric ($m=0$) mode in each multiplet will dominate, leading to a simpler pulsation spectrum.  In very rapidly rotating stars, the oblateness alters the pulsation cavity and further complicates the pattern.  However, for rotation rates less than about 50\% of Keplerian break-up, the radial modes ($l=0$) and the axisymmetric dipolar modes ($l=1$, $m=0$) are still expected\citep{Reese++2008} to follow a regular spacing \black{that is similar to the non-rotating case, but with a slightly smaller~\Dnu}.  

To search for regular patterns we have used observations from the \textit{Transiting Exoplanet Survey Satellite} (\tess), which provides light curves for many thousands of \dsct\ stars at rapid cadence (120-s sampling).
\black{\sout{This gives an order-of-magnitude increase over previously available data.}}
We used the first nine 27-day sectors of \tess\ data and focussed on identifying \dsct\ stars that pulsate at high frequencies (above about 30\,\cd). We also examined stars not previously known to pulsate by calculating the Fourier spectra of \tess\ light curves and measuring the skewness of the distribution of peak heights\cite{Murphy++2019} above 30\,\cd\ as a way to flag likely detections.  

We then inspected the pulsation spectra for regularity using \echelle\ diagrams (described below).  
In addition, we used data from the \kepler\ spacecraft, which observed about 300 \dsct\ stars at short cadence (60-s sampling) during its four-year nominal mission\citep{Balona2015,Bowman+Kurtz2018,Murphy++2019}.  Stars observed in \kepler's long-cadence mode (29.4-min sampling) were not considered because the Nyquist frequency of 24.5\,\cd\ makes it very difficult to identify patterns in high-frequency pulsators.

\begin{figure}
\begin{center}
\ifarxiv
\includegraphics[width=0.6\linewidth]{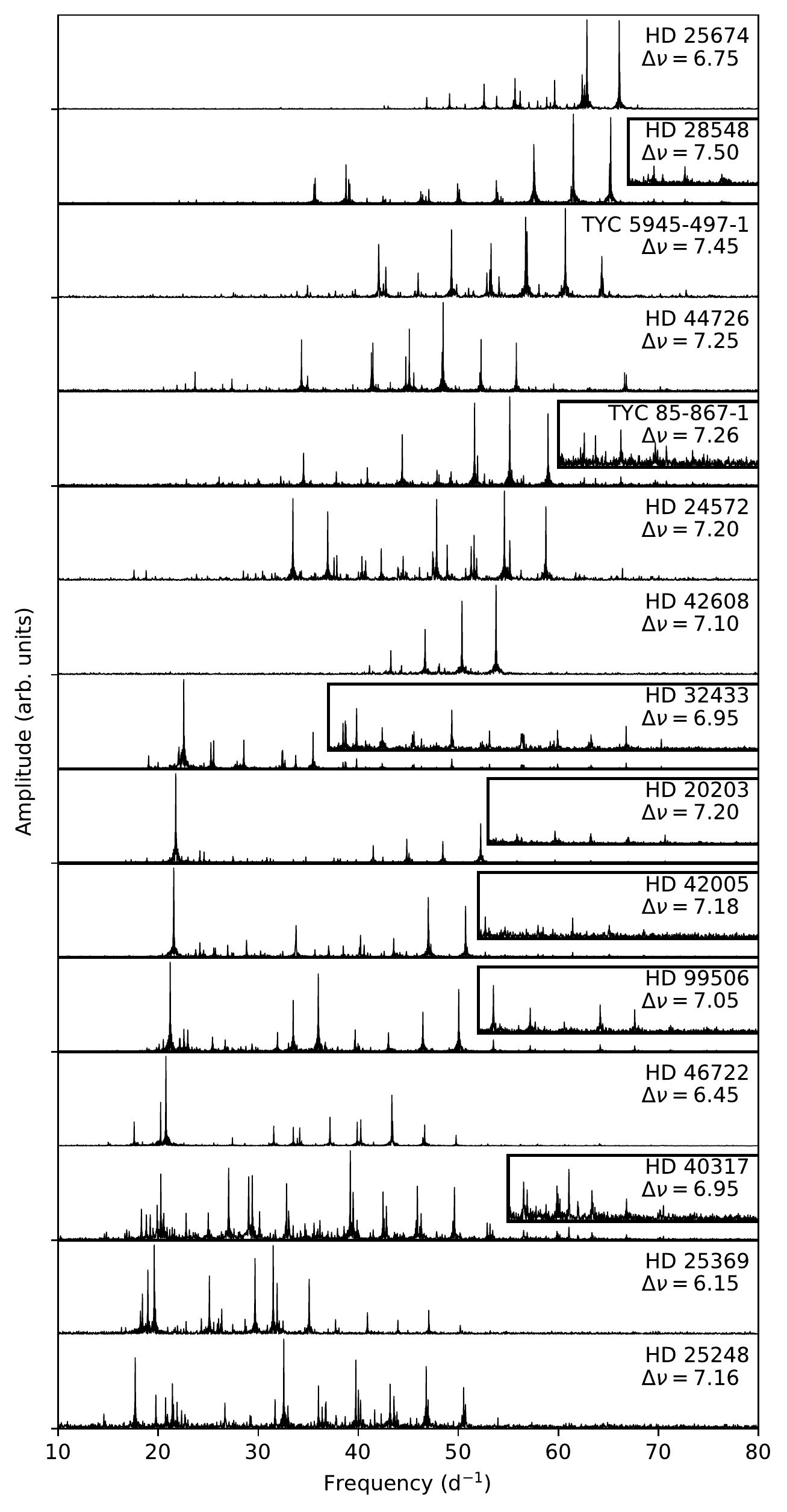}
\fi
\end{center}
\caption{\textbf{Pulsation spectra of 15 high-frequency \dsctbf\ stars observed with \tess. } \black{The measured value of \Dnu\ (in \cd) is given in each panel (see Extended Data Table~\ref{tab:properties}).}  Insets for some spectra expand the vertical axis by a factor of four to make weaker peaks more visible. 
\label{fig:amp-best}}
\end{figure}

We discovered 60 stars with regular frequency spacings (Extended Data Table~\ref{tab:properties}), which define a group of \dsct\ stars for which mode identification is possible.  Fig.~\ref{fig:amp-best} shows some of the pulsation spectra, which have remarkably regular patterns of peaks.  
The small amplitudes of the highest-frequency modes may indicate that turbulent pressure, rather than the standard opacity mechanism, is responsible for driving them\cite{Antoci++2019}.
\black{About one-third of the} stars in our sample (for example, the bottom half of Fig.~\ref{fig:amp-best}) show a strong peak \black{in the range 18--23\,\cd}, which is likely to be the fundamental radial pressure mode ($n=1$, $l=0$).  This identification is strengthened by the fact that these peaks agree with the established period--luminosity relation for the fundamental radial mode in \dsct\ stars\citet{Ziaali++2019}, \black{and by the fact that we find a good correlation between this frequency (when present) and the measured value of \Dnu\ (Extended Data Fig.~\ref{fig:Dnu-fund})}.  In addition, \black{six} stars show a mode that is a factor \black{about $0.78$ shorter in period}, consistent with being the first radial overtone ($n=2$, $l=0$)\citet{Petersen+1996}.  

\begin{figure}
\ifarxiv
\begin{center}
\includegraphics[width=0.94\linewidth]{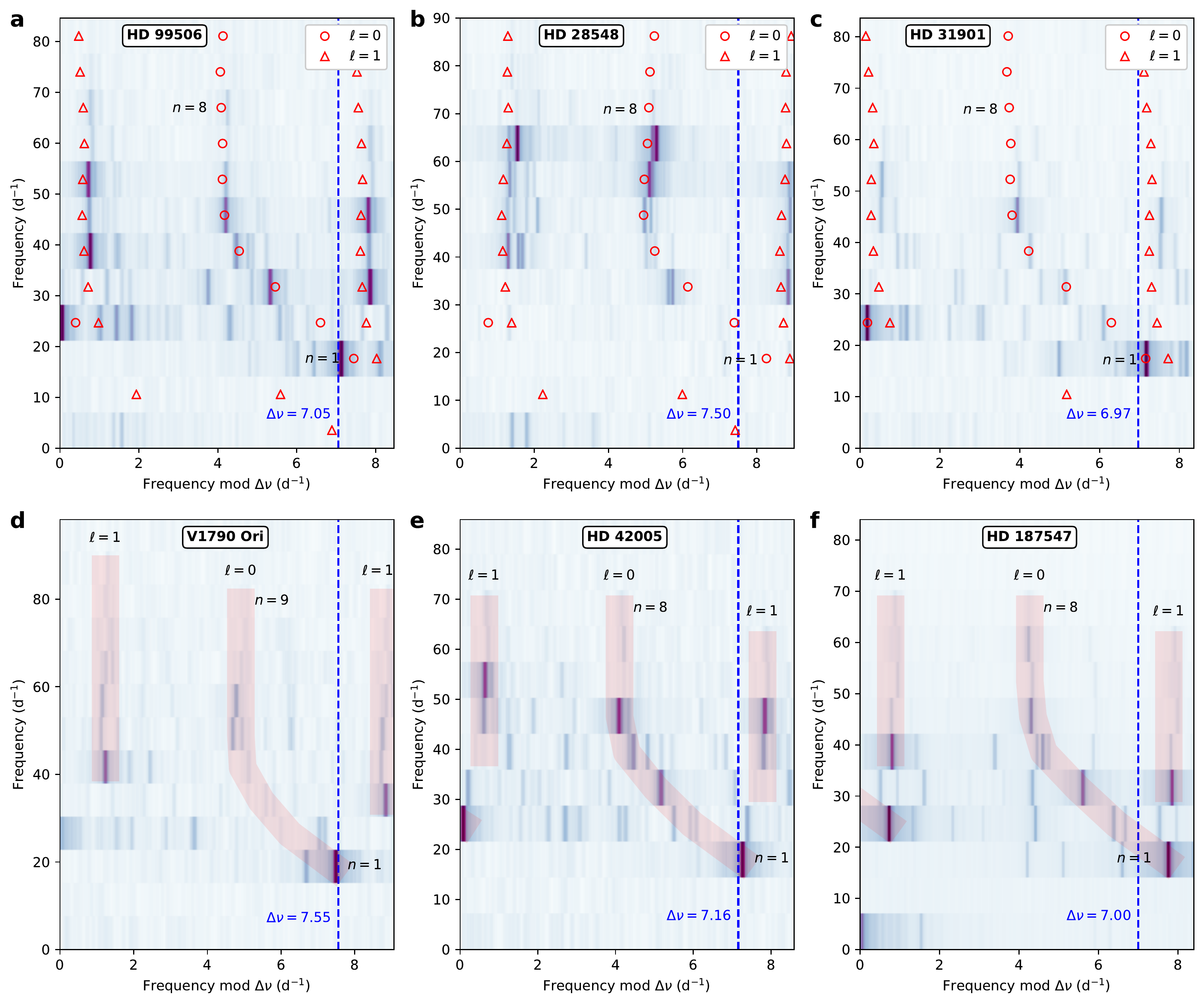}
\end{center}
\fi

    \caption{\textbf{Mode identification in \dsctbf\ stars.} \textbf{a}--\textbf{f}, Pulsation spectra of 
    HD~99506 (\textbf{a}),
    HD~28548 (\textbf{b}),
    HD~31901 (\textbf{c}),
    V1790~Ori (\textbf{d}),
    HD~42005 (\textbf{e}) and
    HD~187547 (\textbf{f}).
    Spectra are shown in \echelle\ format, segments of equal length being stacked vertically.  In each panel, the vertical dashed line shows the value of \Dnu, with a repeated overlap region added on the right for clarity.  The greyscale shows the observed amplitude spectrum, which in most cases was calculated from one 27-day sector of data from the \tess\ spacecraft.  The exception is HD~187547, for which observations were made over 960\,d with the \kepler\ spacecraft\cite{Antoci++2014}.  Some smoothing was applied to the observed amplitude spectra before plotting.  In the top row (\textbf{a}--\textbf{c}), the red symbols show mode frequencies calculated from theoretical models of non-rotating stars, chosen to match the observed modes reasonably well (see Methods).  These allow mode identifications in other stars, as shown in the bottom row (\textbf{d}--\textbf{f}), where the red stripes mark overtone sequences of $l=0$ and $l=1$ modes.  The parameters of the models shown in \textbf{a}--\textbf{c} are as follows (while noting that other values of the parameters also give fits of similar quality): 
\textbf{a,}~HD~99506:
mass $1.68\,\Msun$, metallicity $\mbox{[Fe/H]} = 0.0$, age 200\,Myr, effective temperature 8065\,K and radius 1.51\,\Rsun. 
\textbf{b,}~HD~28548:
mass $1.59\,\Msun$, metallicity $\mbox{[Fe/H]} = -0.2$, age $270$\,Myr, effective temperature 8202\,K and radius 1.41\,\Rsun. 
\textbf{c,}~HD~31901:
mass $1.77\,\Msun$, metallicity $\mbox{[Fe/H]} = 0.08$, age $102$\,Myr, effective temperature 8083\,K and radius 1.51\,\Rsun.
\label{fig:models}}
\end{figure}

\begin{figure}
\ifarxiv
\begin{center}
\includegraphics[width=1.0\linewidth]{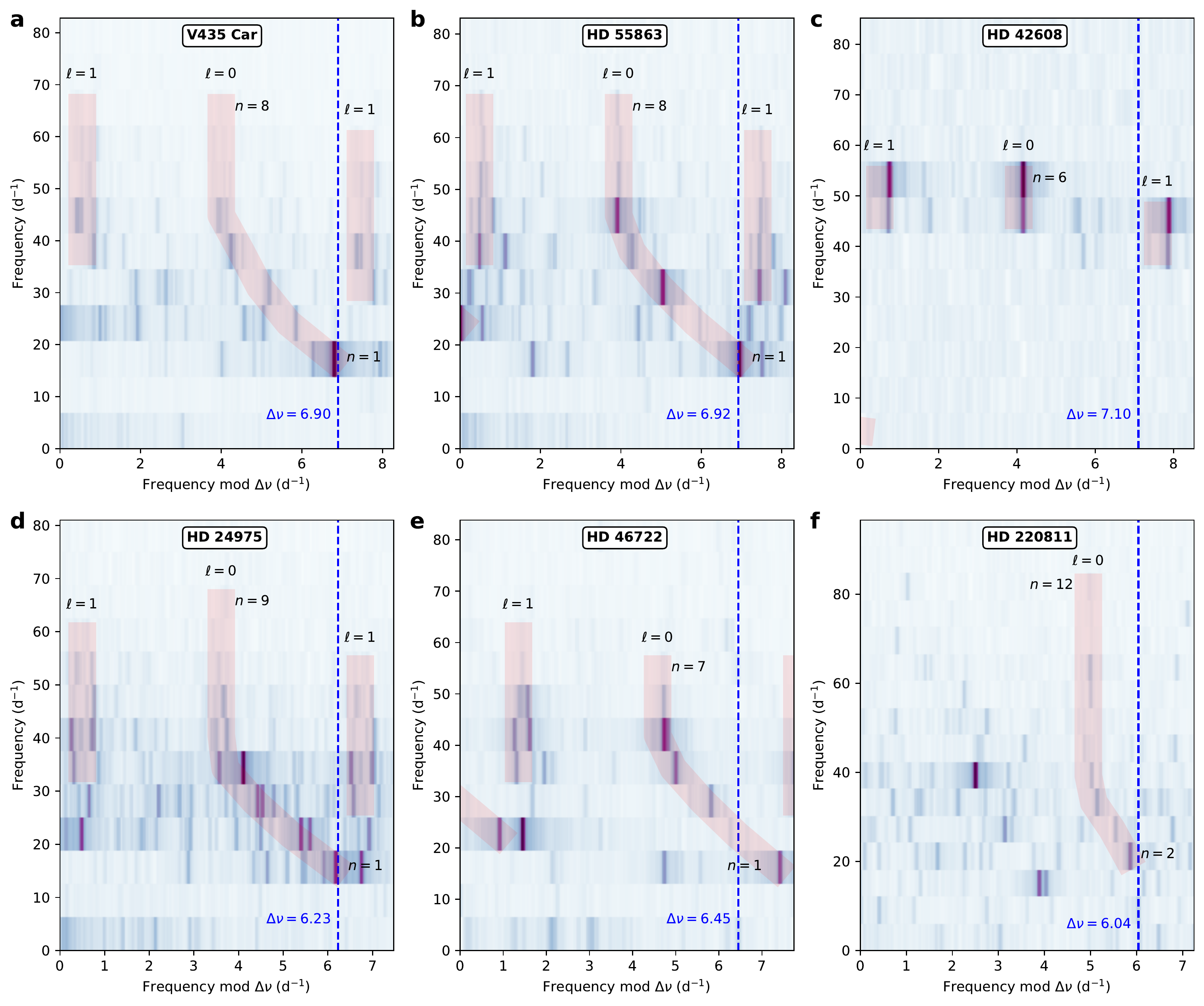}
\end{center}
\fi

\caption{\textbf{More examples of mode identifications in \dsctbf\ stars.} The amplitude spectra are shown in \echelle\ format, with segments of equal length being stacked vertically.  
\textbf{a}, V435 Car; \textbf{b}, HD 55863; \textbf{c}, HD 42608; \textbf{d}, HD 24975; \textbf{e}, HD 46722; \textbf{f}, HD 220811 .
The vertical dashed line shows the value of \Dnu\ used in each case, with a repeated overlap region added on the right for clarity.  The greyscale shows the observed amplitude spectrum of data from the \tess\ spacecraft, where the number of 27-day sectors was four for V435~Car, three for HD~55863, two for HD~24975 and HD~46722, and one for HD~42608 and HD~220811.
Smoothing was applied to the observed amplitude spectra before plotting, and the red stripes mark overtones sequences of $l=0$ and $l=1$ modes. \label{fig:more-echelles} }
\end{figure}

Fig.~\ref{fig:models} shows the pulsation spectra of several \dsct\ stars in \echelle\ format, where the spectra have been divided into equal segments \black{of width $\Delta \nu$ and} stacked vertically so that peaks with the same degree fall along vertical ridges. The regularity of the patterns is striking, similar to \echelle\ diagrams of solar-like oscillators\cite{Aerts++2010-book,Garcia+Ballot2019,Hekker+ChD2017} but at much lower radial orders.  Comparison with pulsation frequencies calculated from theoretical models (red symbols in Fig.~\ref{fig:models}a--c) enables an unambiguous identification of ridges corresponding to sequences of radial modes ($l=0$) and dipolar modes ($l=1$), as shown (more examples are shown in Extended Data Fig.~\ref{fig:more-echelles}).  Sequences of $l=2$ modes do not appear to be present in these stars.

\begin{figure}
\ifarxiv
\begin{center}
\includegraphics[width=\linewidth]{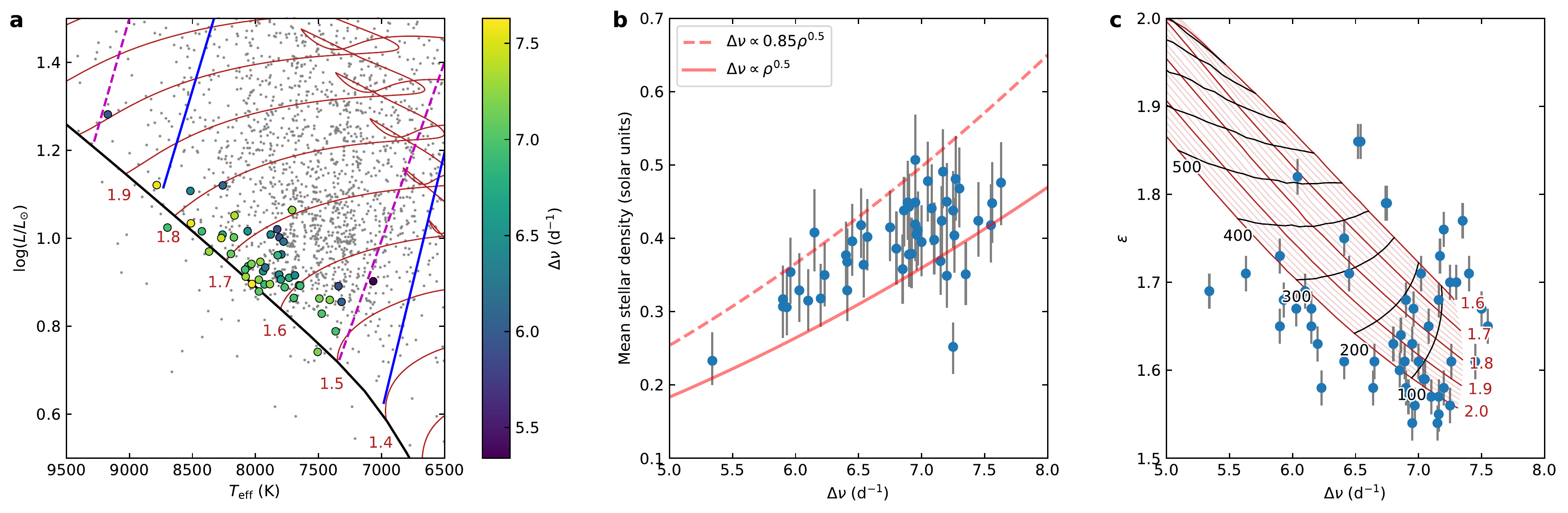}
\end{center}
\fi
\caption{\textbf{Properties of high-frequency \dsctbf\ stars.} \textbf{a,} Location of our sample in the H--R diagram (filled circles, colour-coded by the measured large-frequency separation; $L$, luminosity; \Teff, effective temperature).  The small points show \dsct\ stars observed by the \kepler\ Mission\cite{Murphy++2019} and the red curves (labelled by mass in solar units) are evolutionary tracks calculated for solar metallicity (see Methods).  The solid blue lines show the edges of the theoretical \dsct\ instability strip, and the dashed magenta lines show the observed instability strip based on \kepler\ stars\cite{Murphy++2019}.
\textbf{b,}~Mean stellar density, $\rho$, versus large frequency separation as determined from observations (symbols; error bars, 1$\sigma$ uncertainties), as predicted from the standard scaling relation (solid red line) and from \black{non-rotating} stellar models (red dashed line). 
Stars with close binary companions have been omitted from panels \textbf{a} and \textbf{b} (see Methods).
\black{\textbf{c,}~The phase term~$\varepsilon$, which measures the absolute position of the oscillation spectrum, versus large frequency separation (symbols; error bars, 1$\sigma$ uncertainties).  Red curves (labelled by mass in solar units) are evolutionary tracks based on fitting to radial modes with $n=4$ to~8 (see Methods), and shorter black curves are the corresponding isochrones, labelled in Myr. These models are only intended to be indicative, since they are calculated for solar metallicity and do not include rotation, which affects both \Dnu\ and~$\varepsilon$.  The models do show that, unlike for solar-type stars\cite{White++2011}, $\varepsilon$ varies substantially during the evolution and is therefore sensitive to age, which is an important bonus for asteroseismology of \dsct\ stars.}

\label{fig:HRD}}
\end{figure}

\begin{figure}
\ifarxiv
\begin{center}
\includegraphics[width=0.4\linewidth]{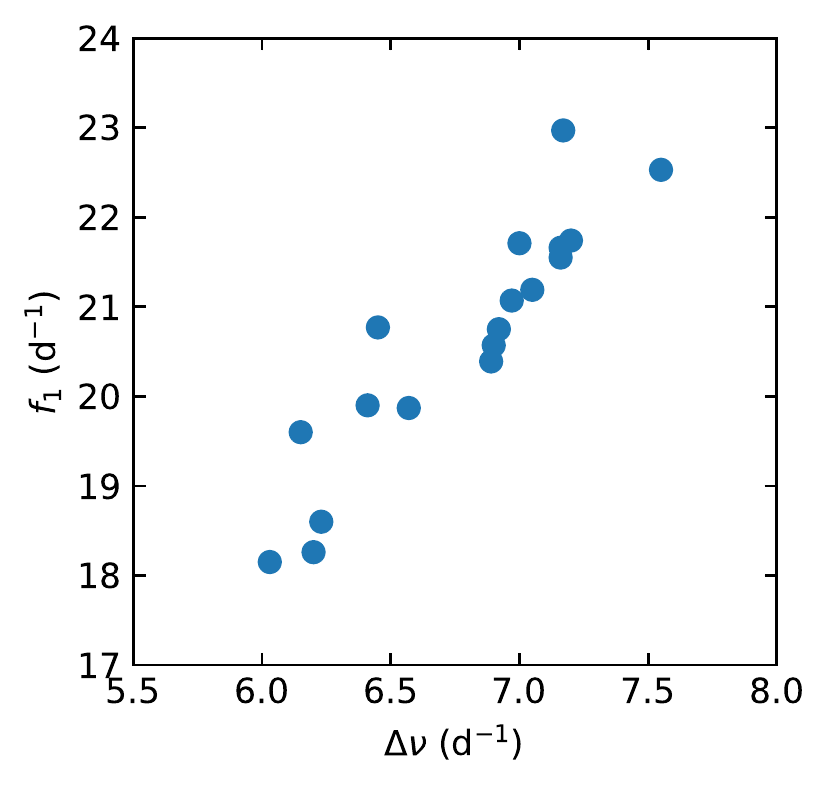}
\end{center}
\fi

\caption{\black{\textbf{Correlation between large separation and the frequency of the fundamental radial mode.}  Symbols show for 18 \dsct\ stars in which the fundamental radial mode ($f_1$) is clearly identified.  A correlation is expected because both quantities depend on the mean stellar density.   We do not expect a perfect correlation due to departures from the asymptotic relation\cite{Aerts++2010-book,Garcia+Ballot2019,Hekker+ChD2017} (see Methods) and variations in~$\varepsilon$ from star to star (see Fig.~\ref{fig:HRD}c).}  \label{fig:Dnu-fund}}
\end{figure}

We have placed our sample in the Hertzsprung--Russell (H--R) diagram using effective temperatures and luminosities derived from broadband colors and \gaia\ parallaxes (Fig.~\ref{fig:HRD}a).  The \dsct\ stars with regular frequency spacings tend to be located near the zero-age main-sequence (ZAMS), with masses between 1.5 and 1.8\,\Msun.  The fact that these stars are relatively young helps to explain their regular pulsation spectra.  In more evolved stars, the nonradial modes are expected to be `bumped' from their regular spacings when they undergo avoided crossings due to coupling with gravity (buoyancy) modes in the core\cite{Hekker+ChD2017}.  For young stars, this mode bumping only occurs at the lowest frequencies, as can be seen from the models of $l=1$ modes in Fig.~\ref{fig:models}a--c (red triangles at low frequencies).

The large frequency separation, \Dnu, scales approximately as the square root of the mean stellar density\cite{Reese++2008,Suarez++2014,Garcia++2015,Paparo++2016-results}.  However, the mode spacings of stars are not completely regular---even in the asymptotic regime---meaning that \Dnu\ varies with frequency.  We used theoretical models to calculate \Dnu\ for \dsct\ stars in the same region that we measured it, namely from radial modes with orders in the range $n=4$ to~8 (see Methods).  We found that \Dnu\ in the models was typically 15\% lower than would be obtained by scaling from the density of the Sun, which is consistent with previous results\cite{Suarez++2014,Garcia++2015,Paparo++2016-results,Mirouh++2019}.  Fig.~\ref{fig:HRD}b compares the observed large separations of our sample with the densities derived from fitting to evolutionary tracks in the H--R diagram.  The results confirm there is a correlation, with most stars lying between the values based on the standard scaling relation (solid red curve) and those from the model calculations (dashed red curve).  Some of the spread is probably due to the range of metallicities of the sample, and some will be due to rotation.  \black{For example, if a star is oblate due to rotation then the mean density will be reduced.}
In addition, the inclination of its rotation axis affects the observed position in the H--R diagram\cite{Suarez++2002} (and hence the inferred radius, mass and density).
\black{The absolute position of the regular comb pattern, parametrised by the phase term $\varepsilon$ (see Methods), also contains important information about the interior structure of the star.  In solar-type stars, the value of~$\varepsilon$ does not change greatly during evolution\cite{White++2011}.  In these intermediate-mass stars, this appears not to be the case and $\varepsilon$ serves as a useful indicator for age (Fig.~\ref{fig:HRD}c).}

High-resolution spectroscopy can be used to measure \vsini, the projected rotational velocity of a star (where~$v$ is the equatorial velocity and~$i$ is the inclination angle), and most intermediate-mass stars have values\cite{Zorec+Royer2012} in the range 50--220\,\kms.  Measurements are available for \black{39 of the 60 stars in} our sample (see Extended Data Table~\ref{tab:vsini}), \black{of which 17 stars have $\vsini \le 50\,\kms$}.  Thus, our sample of \dsct\ stars includes many with unusually low projected rotational velocities, which is consistent with the idea that regular frequency spacings are more common in stars seen at high inclinations (close to pole-on).  

\begin{figure}
\ifarxiv
\begin{center}
\includegraphics[width=0.66\linewidth]{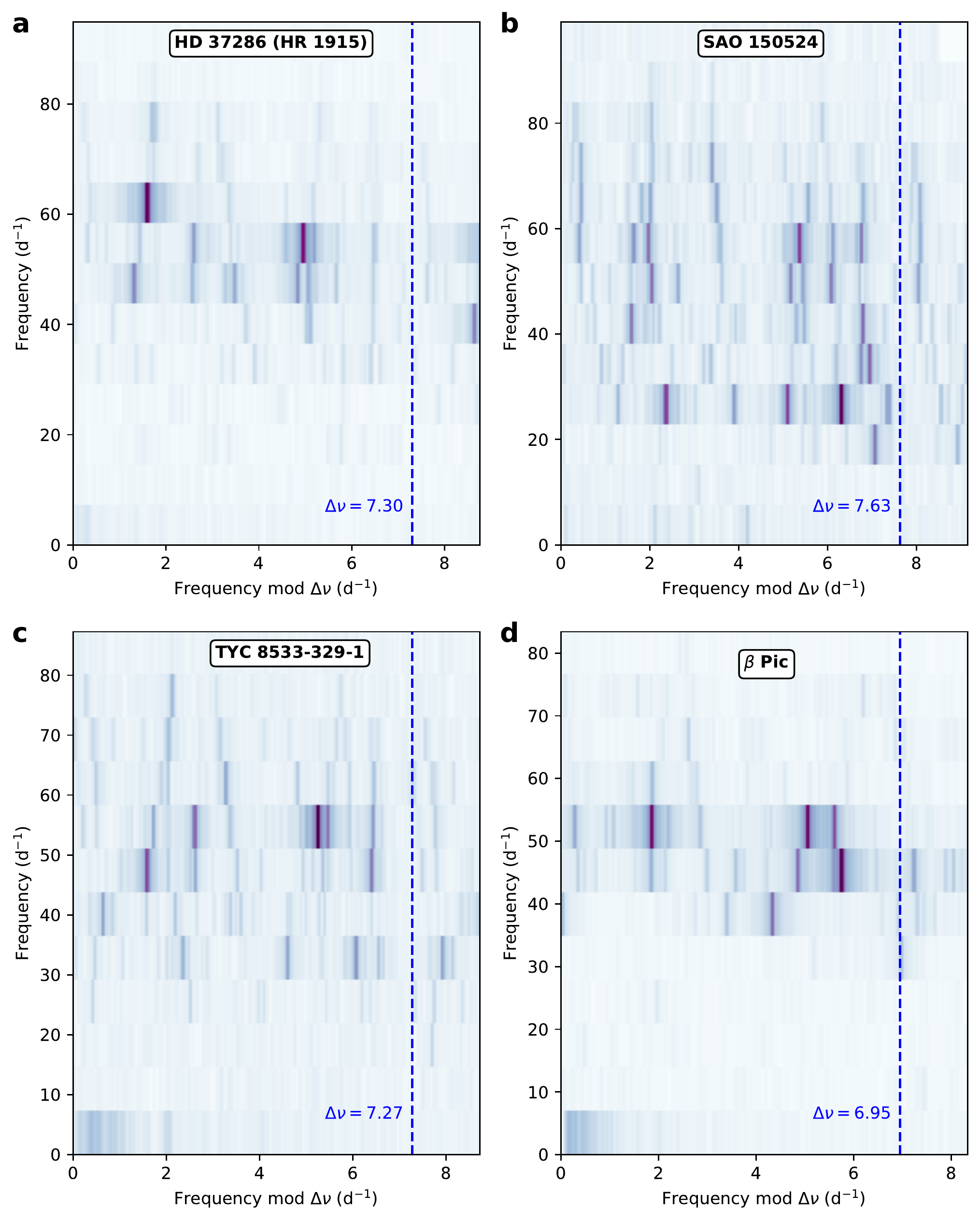}
\end{center}
\fi
\caption{\textbf{Examples of more complicated \echelle\ diagrams of \dsctbf\ pulsations.} \textbf{a}--\textbf{d}, Diagrams for HD~37286 (=HR~1915; \textbf{a}), SAO 150524 (\textbf{b}), TYC 8533-329-1 (\textbf{c}) and $\beta$~Pic (\textbf{d}).  There are sets of ridges at a range of angles, indicating slightly different spacings.  An intermediate value of \Dnu\ was chosen for these diagrams \black{(see Methods)}. 
\label{fig:ridges}}
\end{figure}

Some \echelle\ diagrams show the modes along the $l=1$ ridge to be split into close doublets, as expected for rotating stars (some examples, \black{namely HD~24975 and HD~46722,} are shown in Extended Data Fig.~\ref{fig:more-echelles}).  \black{Four} \echelle\ diagrams show more complicated patterns, with additional ridges at various angles that indicate sequences with slightly different spacings (Fig.~\ref{fig:ridges}).  The rotation axes of these stars are presumably at \black{\sout{lower}higher} inclinations than those with simpler pulsation spectra, which would lead us to expect one $l=0$ ridge and three $l=1$ ridges.  
Beyond the usual rotational splitting of $l=1$ modes, slightly different frequency spacings are expected for each $m$ in an oblate star.  This is because modes with different $m$ propagate along different paths through the star, giving different values for the sound-speed crossing time and hence for~\Dnu.
In stars with even more ridges, the additional sequences could correspond to modes with higher degrees ($l\ge2$)---where coupling between modes with different degree may also be important---and perhaps also to chaotic modes\cite{Reese++2017,Evano++2019}. 

The \black{identification} of regular pulsation frequency patterns in intermediate-mass stars will expand the reach of asteroseismology to new frontiers.  One example is to determine the ages of young moving groups, clusters and stellar streams, which can vary by a factor of up to two, depending on the method used\citep{Mamajek+Bell2014}.  
Spectroscopic radial velocities and \gaia\ astrometry show that several stars in our sample are members of nearby young associations (references given in Extended Data Table~\ref{tab:properties}), including the Octans association (HD~44930, HD~29783, HD~42915), the Carina association (HD~89263), the Columba association (HD~37286 = HR~1915), the $\beta$~Pictoris moving group ($\beta$~Pic itself) and the recently discovered Pisces--Eridanus stellar stream (HD~31901). For the last, gyrochronology yielded an age similar to the Pleiades (about 130\,Myr)\cite{Curtis++2019}, in contrast to the initial approximately 1-Gyr age determination from suspected evolved moving group members\cite{Meingast++2019}.  Asteroseismic modelling of HD~31901 (Fig.~\ref{fig:models}c) clearly confirms a young age for this member of the Pisces--Eridanus group \black{(see Methods)}, and similar age determinations might be possible for other groups containing intermediate-mass stars.

Four stars in our sample (HD~28548, HD~34282, TYC~5945-497-1 and V1790~Ori) exhibit excess emission in the WISE passbands, indicating a circumstellar dust disk.  One of these (HD~34282) has a disk that has been resolved by ALMA, showing it to be inclined $60^\circ \pm 1^\circ$  to the line of sight\citep{vanderPlas2017}.  The constraints on age and inclination of this host star provided by an analysis of its pulsations could illuminate the origin of stellar obliquity\citep{Lai2014} and the pace of disk evolution\citep{Williams2011}.

Finally, we note that six stars in our sample have been classified spectroscopically as \lboo\ stars (references given in \black{Extended Data} Table~\ref{tab:properties}), meaning that their surface chemical abundances show evidence for accretion from circumstellar material.  Given that \lboo\ stars are rare, making up only about 2\% of A stars\cite{Paunzen2001}, the relatively high occurrence rate in our sample lends support to the hypothesis that \lboo\ stars tend to be young, with circumstellar material accreting from a proto-planetary disk. 

\black{The stars observed by \tess\ at 2-min cadence constitute a small fraction of stars that fall on the full-frame images (FFIs).  Future \tess\ observations should reveal many more examples of \dsct\ stars with high-frequency overtones, especially given that the cadence of \tess\ FFIs will switch from 30 minutes to 10 minutes in the extended mission that starts in July 2020.  It is likely that the stars with regular patterns can guide mode identification in the much larger number of \dsct\ stars whose pulsation spectra are not as regular.}

\clearpage


\ifarxiv\clearpage\fi

\ifarxiv\clearpage\fi

\ifarxiv\clearpage\fi

\clearpage


\noindent\textbf{\large Methods}
\medskip

\noindent\textbf{Pulsation analysis} 

\noindent 
Light curves from \tess\citet{Ricker++2015} \black{and \kepler\cite{Borucki++2010}} were downloaded from MAST (Barbara A. Mikulski Archive for Space Telescopes)\cite{MAST}.  We used the Pre-search Data Conditioning Simple Aperture Photometry (PDCSAP) to calculate the Fourier amplitude spectra \black{using a standard Lomb--Scargle periodogram}. 

\black{For \tess, we examined all 92,000 stars having 2-min light curves in sectors 1--9.  We used the skewness of the distribution of peak heights\cite{Murphy++2019} above 30\,\cd\ as a way to identify high-frequency \dsct\ pulsators, producing a list of about 1000 stars.  Inspecting their \echelle\ diagrams (see below) revealed 57 \dsct\ stars having a regular series of high-frequency peaks.  For \kepler, we looked at all (about 330) \dsct\ stars that have short-cadence data (60-s sampling) and identified three stars with regular peaks.}

The large separations (\Dnu) for the 60 stars in our sample are listed in Extended Data Table~\ref{tab:properties}.  In most cases, \Dnu\ was measured by aligning the highest-frequency radial modes in a vertical ridge in the \echelle\ diagram \black{using the Python package \texttt{echelle}\cite{Hey2019}, which allows the value of \Dnu\ to be fine-tuned interactively.  This allowed \Dnu\ to be measured to a precision of about 0.02\,\cd}\ (see examples in Fig.~\ref{fig:models} and Extended Data Fig.~\ref{fig:more-echelles}).  
\black{Four stars do not show a clear sequence of radial modes, with the \echelle\ diagrams showing several ridges that are not quite parallel (Fig.~\ref{fig:ridges}).  In these cases, we chose \Dnu\ to be the average of the values needed to make the individual ridges vertical.}

The phase term $\varepsilon$ is given for those stars having a clear $l=0$ sequence, as determined from the horizontal position of that ridge in the \echelle\ diagram.  Note that \Dnu\ and~$\varepsilon$ are related to the frequencies of high-order radial modes via the asymptotic relation\cite{Aerts++2010-book,Garcia+Ballot2019,Hekker+ChD2017}: $\nu_{n,l=0} \approx \Delta\nu (n + \varepsilon)$. \black{The uncertainty in $\varepsilon$ determined in this way is about 0.02.}

To rule out contamination from nearby stars as the source of the observed pulsations, we examined the pixel data and cross-matched with the \gaia\ DR2 catalogue. \black{We considered a region of 5$\times$5 TESS pixels (63$\times$63\,arcsec) centred on each target.  We found that no dilution is present in one-third of the targets, with most of the remainder having small amounts of dilution (0.1--3\%).  Only five stars have dilutions above 8\%.  We conclude that contamination of the photometry from nearby stars is negligible.  }

\bigskip
\noindent\textbf{Fundamental stellar properties}

\noindent
To estimate properties for our sample we used Tycho $B_{T}$ and $V_{T}$ photometry\cite{hog00}, which we transformed into Johnson $B$ and $V$ magnitudes\cite{bessell00}. We then used a ($B-V$)--\Teff\ relation\cite{flower96}, \gaia\ DR2 parallaxes\cite{lindegren18}, a 3D dust map\cite{bovy16}, and $V$-band bolometric corrections to calculate effective temperatures and luminosities.  We did this by solving for the distance modulus, as implemented in the `direct mode' version of \texttt{isoclassify}\cite{huber++2016}. For stars with typical uncertainties $> 0.01$\,mag in Tycho ($V_{T} > 9$\,mag), we used the \gaia\ $BP-RP$ colour index (with which we interpolated the colour--\Teff\ relation in the MIST (MESA Isochrones and Stellar Tracks) model grid\cite{choi16} for solar metallicity) to derive \Teff, and we used 2MASS K-band magnitudes in combination with \gaia\ parallaxes to derive luminosities. 
 
We adopted 2\% fractional uncertainties for all derived effective temperatures, which is typical of the residual scatter in optical colour--temperature relations\cite{casagrande11}. A comparison of our \gaia-derived temperatures with those derived from Tycho photometry for stars with $V_{T} < 10$\,mag, and a comparison with an independent implementation of the infrared flux method (IRFM), both showed good agreement with no significant systematic offsets. Our effective temperatures are on average about 1.5\% (200\,K) hotter than those for A-type stars in the \kepler\ Stellar Properties Catalog\cite{Huber2014, Mathur2017}, which were predominantly based on the Kepler Input Catalog (KIC)\cite{Brown2011}. Such systematic differences are typical for effective temperature scales in A stars, reflecting the fact that the KIC was not optimized for A stars.

To estimate mean stellar densities, we fitted the effective temperatures and luminosities derived in the previous step to MIST isochrones using the `grid mode' of \texttt{isoclassify}, assuming a solar-neighborhood metallicity prior. The procedure also yielded estimates of stellar masses and surface gravities, which combined with T$_{\rm{eff}}$ were used for the interpolation of bolometric corrections in the previous step. We iterated between the `direct mode' and `grid mode' calculations until all values converged\black{, and adopted 0.03\,mag bandpass-independent uncertainties in reddening and bolometric corrections}.  Extended Data Table~\ref{tab:properties} lists all stellar properties of the sample. Typical uncertainties are about \black{5}\,\% in luminosity and about \black{15}\,\% in mean stellar density.
The properties of V1366~Ori (HD~34282) are not shown because they are highly uncertain due to obscuration by circumstellar material (it is classified as a Herbig Ae star)\cite{Casey++2013-HD34282}.  This star is not plotted in Fig.~\ref{fig:HRD}.

To identify close binaries, which could bias the derived stellar parameters, we cross-matched our targets with the Washington Double Star catalogue (WDS).  We also calculated the \gaia\ DR2 re-normalized unit weight error (RUWE) for each target, which provides a quality metric that accounts for the effects of colour and apparent magnitude on \gaia\ astrometric solutions.  Stars with WDS companions within 2\,arcsec or \gaia\ RUWE $>$ 2 do not have parameters in Extended Data Table~\ref{tab:properties} and were not plotted in Fig.~\ref{fig:HRD}.

\bigskip
\noindent\textbf{High-resolution spectroscopy}

\noindent
We obtained optical high-dispersion spectra of some stars in the sample in April and May 2019 using the HIRES spectrograph\cite{vogt94} at the Keck-I 10-m telescope on Maunakea observatory, Hawai`i. The spectra were obtained and reduced as part of the California Planet Search queue\cite{howard10}. We typically obtained 1-min integrations using the C5 decker, resulting in a signal to noise ($S/N$) per pixel of 50 at 600\,nm with a spectral resolution of $R\approx 60,000$. 

High-resolution spectra for some stars were obtained in May and June 2019 using the NRES spectrograph\cite{siverd2018} at the Las Cumbres Observatory Global Telescope Network\cite{brown2013} 1-m telescopes at Cerro Tololo Inter-American Observatory, Chile, and at Sutherland, South Africa. Exposure times were typically 10 min, resulting in a $S/N$ per resolution element above 70 at about 510\,nm, with a spectral resolution of $R\approx 50,000$.
\black{High-resolution spectra for an additional nine stars were obtained in June 2019 using the Veloce Rosso spectrograph\cite{Gilbert++2018} at the 3.9-m Anglo-Australian Telescope (AAT).  These spectra covered the range 580--930\,nm at a resolution of $R\approx 75,000$. Typical exposure times were 5--10 min (in cloudy conditions), resulting in a $S/N$ per pixel of 50--90 at about 780\,nm.}

\begin{figure}
\ifarxiv
\begin{center}
\includegraphics[width=1.0\linewidth]{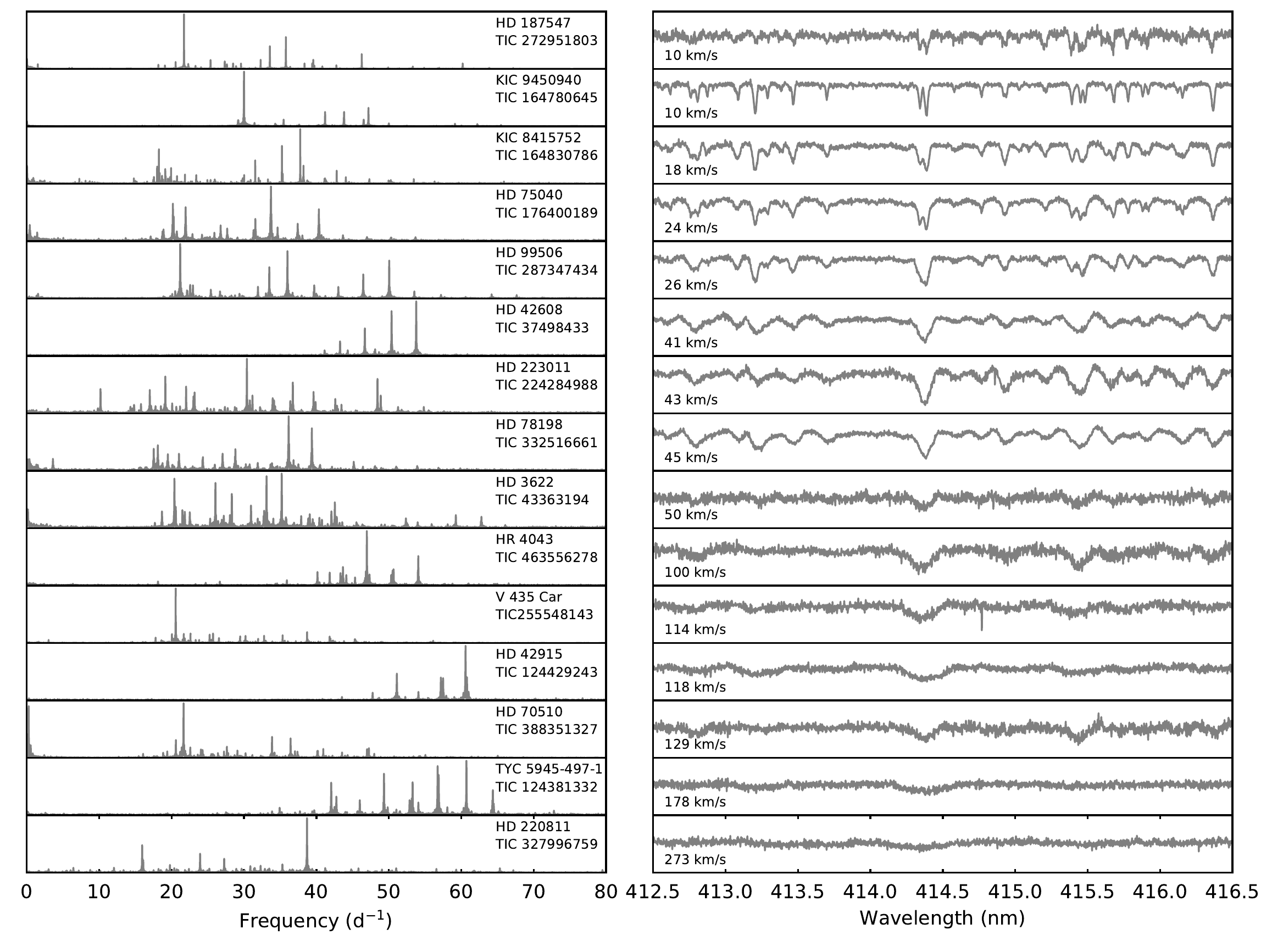}
\end{center}
\fi

\caption{\textbf{Fourier amplitude spectra and high-resolution spectra of high-frequency \dsct\ stars.} 
Left panel, Fourier amplitude spectra of 15 stars; each star has two catalogue names, as shown. Right panel, high-resolution spectra of the stars; measured \vsini\ values are given. The stars are sorted by increasing \vsini\ from top to bottom. See Methods section `High-resolution spectroscopy' for details.
\label{fig:spectra}}
\end{figure}

Extended Data Figure~\ref{fig:spectra} shows a small region of some of these spectra, alongside the Fourier amplitude spectra.
The spectral analysis was performed using the {\sc uclsyn} spectral synthesis package\cite{smith88,smith92} using {\sc atlas9} models without convective overshooting\cite{castelli97}. Atomic data used in the analysis was obtained from the VALD database\citet{kupka99}, using their default search and extraction parameters. Surface gravities were fixed to $\log g = 4.0$ for all stars in the analysis. A microturbulent velocity of  $\xi=3$\,\kms\ was assumed, which is the typical value for stars within the spectral range considered here\citep{niemczura15,niemczura17}.  Measurements of the projected equatorial rotation velocity (\vsini) were obtained through individual fits to several small (5\,nm) regions between 500\,nm and 550\,nm (and 620--650\,nm plus 778\,nm for the AAT spectra), avoiding any inter-order gaps. The final values were determined by calculating the mean and standard deviation of the values obtained in the small spectral regions.

Independent \vsini\ values were determined for \black{five of the} spectra using the Grid Search in Stellar Parameters (\textsc{GSSP}) software\citep{Tkachenko2015}. \textsc{GSSP} is designed to fit a grid of synthetic spectra with varying $T_{\rm eff}$, $\log\,g$, $\xi$, \vsini\ and $[M/H]$ to each observed spectrum and output the $\chi^{2}$ values of the fit. These synthetic spectra are generated on-the-fly during the fitting process using the \textsc{SynthV} radiative transfer code\citep{Tsymbal1996} combined with a grid of atmospheric models from the \textsc{LLmodels} code\citep{Shulyak2004}. We fixed the microturbulent velocity at $\xi=2.0$\,\kms\ to prevent degeneracies with metallicity. The derived values were found to agree \black{within uncertainties} with the results from the {\sc uclsyn} spectral synthesis.

\black{For a further nine stars, we estimated \vsini\ using low-resolution spectra that were obtained either with the RSS instrument on the Southern African Large Telescope (SALT)\citep{Burgh+2003,Kobulnicky+2003,Buckley+2006} or the ISIS instrument on the William Herschel Telescope (WHT). Exposure times were typically a few minutes, which provided a $S/N$ of about $100$ at a spectral resolution of $R\approx 3,000$. For each target, a coarse grid of synthetic models was constructed using the stellar parameters in Extended Data Table~\ref{tab:properties} and a range of \vsini\ values. The observations were compared to the synthetic spectra to estimate the \vsini\ and the associated uncertainty.}

Extended Data Table~\ref{tab:vsini} lists the determined \vsini\ values for each star.  Values in parentheses indicate close binaries (see above), meaning that \vsini\ may not be reliable. \black{\sout{We included six stars (below the line) for which we obtained spectroscopy but which are not in Extended Data Table~\ref{tab:properties} because their pulsations---while at high frequency---do not show a clear value of \Dnu.}}

To determine membership of moving groups, clusters and stellar streams, we calculated barycentric radial velocities using the Python implementation {\tt barycorrPy}\citep{Kanodia2018} of the barycentric correction algorithm of Wright et al.\cite{Wright2014}.  These were combined with space motions calculated from \gaia\ DR2 astrometry, and Bayesian posterior probabilities of membership in known nearby moving groups were calculated using {\tt Banyan $\Sigma$}\citep{Gagne2018}.

\bigskip
\noindent\textbf{Stellar Models} 

\noindent
The stellar models presented in Fig.~\ref{fig:models} used the `astero' extension of MESA (Modules for Experiments in Stellar Astrophysics)\cite{Paxton++2011,Paxton++2013,Paxton++2015}.  We used two approaches that gave similar results. One was based on a model grid calculated with MESA (v8118), where we varied mass from 1.3\,\Msun\ to 1.9\,\Msun\ in steps of 0.01\,\Msun\ and metallicity (\FeH) from $-0.5$ to 0.5 in steps of 0.1.  We used a fixed (solar-calibrated) mixing-length parameter of $\alpha_{\rm MLT}$ = 1.9 and a helium-to-heavy-element enrichment ratio of 1.33. The best-fitting model was found by maximum likelihood estimation, where we included effective temperature, metallicity, luminosity and all identified pulsation frequencies.  
Equal weight was given in the likelihood function to the following five observables: frequencies of radial modes, frequencies of dipolar modes, effective temperature, metallicity and luminosity. 
The other approach used the automated simplex search in MESA-astero (v7503), where the fit was guided by the observed radial modes only. The search was allowed to vary the mass, metallicity, mixing length, and the age of the model in order to converge to the best fit.  A helium-to-heavy-element enrichment ratio of 1.4 was used. Both approaches assumed a primordial helium abundance of 0.249 and we did not make any correction for surface effects in the way that is commonly done for solar-like stars\cite{Ball+Gizon2014}.

\black{For the three examples shown in the upper row of Fig.~\ref{fig:models}, the agreement between models and observations is sufficiently good that we can unambiguously identify the two sequences corresponding to $l=0$ and $l=1$ modes.  One noteworthy feature of the models and the observations is that the $l=0$ sequence bends to the right at the bottom of each figure, indicating that \Dnu\ decreases towards the lowest-order modes, whereas the $l=1$ sequence does not show this effect.  This difference is a general feature of these models and makes it possible to identify the sequences in other stars, as shown in the lower row of Fig.~\ref{fig:models} and in Extended Data Fig.~\ref{fig:more-echelles}.}

For Fig.~\ref{fig:HRD}a we used the evolutionary tracks with solar metallicity ($X = 0.71$, $Z = 0.014$) from Murphy et al.\cite{Murphy++2019}. The other parameters of those tracks are $\alpha_{\rm MLT} = 1.8$, exponential core overshooting of 0.015\,$H_p$ (pressure scale heights), exponential over- and undershooting of 0.015\,$H_p$ for the hydrogen-burning shell, exponential envelope overshooting of 0.025\,$H_p$, diffusive mixing $\log (D_{\rm mix}) = 0$ (with $D_{\rm mix}$ in cm$^2$s$^{-1}$), OPAL opacities and the solar abundance mixture\citet{Asplund2009}. As noted by Murphy et al.\cite{Murphy++2019}, these tracks are in good agreement with the MIST tracks computed with no rotation and similar metallicities, except that the latter have a shorter main-sequence phase. This is not expected to be important for our targets, which are mostly young (close to the ZAMS).  

Although it is possible for \dsct\ pulsations to occur in the pre-main-sequence (PMS) phase, prior to the onset of hydrogen burning\cite{Zwintz++2014}, there is no indication of a PMS classification \black{in the literature} for most of the stars in our sample.  

\bigskip
\noindent\textbf{\black{Detailed modelling of HD~31901.}}

\noindent
\black{As a member of the Pisces--Eridanus stellar stream, this star makes a good test case.  We used the models described above, constrained by the observed frequencies of the radial and dipole modes and by the observed effective temperature and luminosity. Following Curtis et al.\cite{Curtis++2019}, we assumed the metallicity is close to solar.  The results imply a mass of $(1.71 \pm 0.05)$\Msun, a radius of $(1.54 \pm 0.03)$\Rsun\ and an age of $150 \pm 100$\,Myr.  The latter is consistent with the age of about 130\,Myr from Curtis et al.\cite{Curtis++2019} but not with the value of about 1\,Gyr determined by Meingast et al.\cite{Meingast++2019}.  }

\bigskip
\noindent\textbf{Additional references and notes} 

\noindent
As mentioned in the main text, several previous studies have reported regular frequency spacings in the Fourier amplitude spectra of \dsct\ stars\citet{Handler++2000,Garcia++2009,Breger++2009,Breger++2011,Antoci++2011,Zwintz++2011,Zwintz++2013,Paparo++2013,Casey++2013-HD34282,Garcia++2013,Suarez++2014,Maceroni++2014,Garcia++2015,Paparo++2016-method,Paparo++2016-results,Michel2017,Bowman+Kurtz2018}.  Among these, the following stars are included in our sample:
\begin{itemize}

\item HD~187547 (KIC~7548479): the large frequency spacing was previously reported as 40.5\,\muhz\ (3.5\,\cd)\cite{Antoci++2011,Antoci++2014}, which is factor of two smaller than the value we have identified from the same \kepler\ observations. \black{Comparing the \echelle\ diagram of this star (Fig.~\ref{fig:models}) with others in our sample indicates that the larger \Dnu\ is correct.  This is also consistent with the \gaia\ DR2 parallax ($6.57 \pm 0.24$\,mas), which places this star close to the ZAMS.}

\item HD~34282 (V1366 Ori): based on observations with MOST (Microvariability and Oscillations of Stars), Casey et al.\citet{Casey++2013-HD34282} reported a large separation of 3.75\,\cd, which is half the value reported here.  \black{Both values would be consistent with the \hipparcos\ parallax ($5.24 \pm 1.67$\,mas), as used by Casey et al., but the much more precise \gaia\ DR2 parallax ($3.08 \pm 0.29$\,mas) and comparison with other stars in our sample confirms that the larger \Dnu\ value is correct. }  V1366~Ori is a Herbig Ae star\cite{Mora++2001}, so it may be pre-main-sequence.  Its classification in SIMBAD as an eclipsing binary appears to be incorrect.

\item \bpic: known to be a high-frequency \dsct\ star\citet{Mekarnia2017,Zwintz++2019}, but a value for the large separation has not been reported. The TESS observations indicate a value of $\Delta\nu=6.95\,\cd$ (Fig.~\ref{fig:ridges}). 

\end{itemize}

The following stars are not in our sample but seem likely to be high-frequency \dsct\ stars with regular spacings:
\begin{itemize}

\item HD~144277: based on data from MOST and CoRoT (COnvection, ROtation and planetary Transits), Zwintz et al.\cite{Zwintz++2011} suggested a large separation of 7.2\,\cd.  This star will not be observed by \tess\ in its nominal two-year mission\cite{TESS-viewing-tool}, but is scheduled to be observed in sector~39.

\item HD~261711: based on MOST and CoRoT data, Zwintz et al.\cite{Zwintz++2013} suggested a large separation of 6.72\,\cd.  This star was observed by \tess\ in sector~6, but only with 30-minute sampling.  

\item HD~174966: based on CoRoT data, Garc{\'\i}a Hern{\'a}ndez et al.\cite{Garcia++2013} suggested a large separation of 5.53\,\cd.  This star will not be observed by \tess\ in its nominal two-year mission\cite{TESS-viewing-tool}.

\item XX Pyx: based on ground-based multisite observations, Handler et al.\cite{Handler++2000} reported 22 pulsation frequencies in the range 27--76\,\cd and suggested a large separation of 4.63\,\cd.  We have examined the published frequencies for this star using \echelle\ diagrams and confirm that a value of $\Dnu = 4.70\,\cd$ gives a reasonably good alignment of the peaks.  This star will not be observed by \tess\ in its nominal two-year mission\cite{TESS-viewing-tool}, but is scheduled to be observed in sector~35.

\item\black{HD~156623: based on observations with the bRing robotic observatory network, Mellon et al.\cite{Mellon++2019} found frequencies in the range 60--70\,\cd\ and suggested regularity at three different separations: 3.75, 7.25, and 2.75\cd.  This star was observed by TESS in sector~12 and shows a pattern similar to other stars in our sample, with a spacing of $\Dnu=7.31$\,\cd.}

\item\black{HD~27462 (TT~Ret): based on \tess\ data, Khalack et al.\cite{Khalack++2019} preferred a large separation of 3.3\,\cd.  Our examination of the \tess\ data and a comparison with the stars in our sample suggests $\Dnu=6.9$\,\cd.} \black{The WDS catalogue\cite{Mason2001} lists this star as a binary with a separation of 0.4\,arcsec and a magnitude difference of 0.7.  This is consistent with \gaia\ DR2, which gives no parallax and a large astrometric excess noise (RUWE $\approx$ 77).  Accounting for the binary, the \hipparcos\ parallax places the two components close to the ZAMS, consistent with our suggested value of \Dnu. }
\end{itemize}



\bigskip
\noindent\textbf{Data availability}

\noindent
\tess\ and \kepler\ data are available from the MAST portal at \\https://archive.stsci.edu/access-mast-data. All other data are available from the corresponding author upon reasonable request.

\bigskip
\noindent\textbf{Code availability}

\noindent
We have made use of standard data analysis tools in Python, as noted and referenced in Methods.

\bigskip
\bigskip
\bigskip
\bigskip


{
\small\sffamily

}

{\small\sffamily

\paragraph{Acknowledgements:} We gratefully acknowledge the \tess\ and \kepler\ teams, whose efforts made these results possible.
This research was partially conducted during the Exostar19 program at the Kavli Institute for Theoretical Physics at UC Santa Barbara, which was supported in part by the National Science Foundation under Grant No. NSF PHY-1748958.  We thank colleagues in that program, especially Rich Townsend, for many stimulating discussions.  We also thank Andr{\'es Moya, Antonio Garc{\'i}a Hern{\'a}ndez, Juan Carlos Su{\'a}rez and Zhao Guo for comments on the manuscript.}   We gratefully acknowledge support from the Australian Research Council (grant DE~180101104), and from the Danish National Research Foundation (Grant DNRF106) through its funding for the Stellar Astrophysics Center (SAC). 
\blue{D.H. acknowledges support from the Alfred P. Sloan Foundation, the National Aeronautics and Space Administration (80NSSC18K1585, 80NSSC19K0379), and the National Science Foundation (AST-1717000).}
\blue{H.K. acknowledges support from the European Social Fund via the Lithuanian Science Council (LMTLT) grant  No. 09.3.3-LMT-K-712-01-0103.}
Y.L. acknowledges the Joint Research Fund in Astronomy (U1631236) under cooperative agreement between the National Natural Science Foundation of China (NSFC) and Chinese Academy of Sciences (CAS). \black{D.L.H. acknowledges support by the Science and Technology Facilities Council under grant ST/M000877/1.}
The research leading to these results has (partially) received funding from the Research Foundation Flanders (FWO) under grant agreement G0H5416N (ERC Runner Up Project).
This work makes use of observations from the LCOGT network.
\black{This work has made use of data from the European Space Agency (ESA) mission \gaia\ (\texttt{\small https://www.cosmos.esa.int/gaia}), processed by the \gaia\ Data Processing and Analysis Consortium (DPAC,\\
\texttt{\small https://www.cosmos.esa.int/web/gaia/dpac/consortium}). } \black{Some of the observations reported in this paper were obtained with the Southern African Large Telescope (SALT) under programs 2015-2-SCI-007, 2016-2-SCI-015 and 2017-2-SCI-010. The ISIS instrument is mounted on the WHT, which is operated on the island of La Palma by the Isaac Newton Group of Telescopes in the Spanish Observatorio del Roque de los Muchachos of the Instituto de Astrofísica de Canarias.}
\black{The Veloce Rosso facility was funded by an Australian Research Council (ARC) Linkage Infrastructure, Equipment and Facility (LIEF) grants LE150100087 \&\ LE160100014, and UNSW Research Infrastructure Scheme grant RG163088. CGT and CB acknowledge the support of ARC Discovery grant DP170103491.}
The authors wish to recognize and acknowledge the very significant cultural role and reverence that the summit of Mauna Kea has always had within the indigenous Hawaiian community; we are most fortunate to have the opportunity to conduct observations from this mountain.
We also acknowledge the traditional owners of the land on which the Anglo-Australian Telescope stands, the Gamilaraay people, and pay our respects to elders past, present, and emerging.

\paragraph{Author Contributions:}
T.R.B, S.J.M., D.R. Hey, W.J.C., G.L., Y.L., I.L.C. and J.Y. analysed the photometric observations;
T.L., D.S., \black{W.H.B,} T.R.W., D.R.R., J.F. and J.J.H. calculated and/or interpreted theoretical models;
V.A. \black{and H.K.} coordinated the selection of the targets for the \tess\ observations;
D.H., D.R. Harbeck, S.S., B.S., T.M.B., A.W.H., H.I., C.M., M.R., \black{C.B., A.D.R., C.G.T, M.J.I and D.L.H} obtained and/or analysed the spectroscopic observations;
E.G. and A.W.M. identified objects that belong to moving groups; 
\black{G.R.R., R.K.V. and J.M.J. were key architects of the \tess\ Mission.} 
All authors reviewed the manuscript.

\paragraph{Competing Interests}
The authors declare no competing interests. 

\paragraph{Correspondence and requests for materials} should be addressed to T.R.B.
\paragraph{Reprints and permissions information} is available at www.nature.com/reprints.
}

\clearpage
\extended

\ifarxiv\clearpage\fi

\ifarxiv\clearpage\fi

\clearpage
\begin{table}
\small
\caption{\sffamily\textbf{Properties of high-frequency \dsct\ stars}\label{tab:properties}}
\smallskip
\ifarxiv
\\[1ex]
{\sffamily\small HD and TIC indicate star catalogues; $\sf V$, apparent magnitude; \sfTeff, effective temperature (fractional uncertainties are 2\%); $\sf L$, luminosity; $\sf\rho$, mean density; \Dnu, large frequency separation; $\varepsilon$, phase term. Stars without values for \sfTeff, $\sf L$ and $\sf\rho$ are close binaries, meaning that those parameters cannot be reliably calculated (see Methods for details).  References indicate classifications as \dsct\ stars \cite{Holdsworth++2014,Rodriguez2000,Amado++2004,Antoci++2011,Antoci++2014,Casey++2013-HD34282,Mekarnia2017,Zwintz++2019}, \lboo\ stars\cite{Gray++2017,Murphy++2015} and members of young moving groups, clusters or stellar streams\cite{Zuckerman++2011,Torres2006,Mamajek+Bell2014,Murphy+Lawson2015,Paunzen++2015,Meingast++2019}.}
\\[1ex]

\tablehead{\multicolumn{1}{c}{HD} & \multicolumn{1}{c}{Name} & \multicolumn{1}{c}{TIC} &
  \multicolumn{1}{c}{$\sf\it V$} & \multicolumn{1}{c}{\sfTeff} &
  \multicolumn{1}{c}{$\sf L$} & \multicolumn{1}{c}{$\rho$} & \multicolumn{1}{c}{\Dnu} & \multicolumn{1}{c}{$\varepsilon$} & \multicolumn{1}{c}{Refs.} \\
& & & & \multicolumn{1}{c}{(K)} & \multicolumn{1}{c}{(\sfLsun)} &\multicolumn{1}{c}{(\rhosun)} & \multicolumn{1}{c}{(\sfcd)} & \\ \hline
\noalign{\smallskip}
}

\sffamily\footnotesize
\begin{supertabular}{rrrrcrcccl}
  2280 &               & 281499618   &    9.13 & 7510  & 5.52 $\pm$0.26 & 0.49$\pm$0.06 &    7.17 &  1.73 \\
  3622 &               &  43363194   &    7.77 & 7930  & 7.86 $\pm$0.35 & 0.45$\pm$0.06 &    6.89 &  1.61 \\
 10779 &               & 229139161   &    8.78 & 7730  & 8.13 $\pm$0.36 & 0.39$\pm$0.05 &    6.80 &  1.63 \\
 10961 &               & 231014033   &    9.39 & ---   & \emdash        & \emdash       &    7.30 &  1.70 \\
 17341 &               & 122615966   &    9.32 & 7810  & 10.05$\pm$0.50 & 0.32$\pm$0.05 &    5.90 &  1.73 &\cite{Gray++2017}\\
 17693 &               & 122686610   &    7.80 & 7880  & 10.21$\pm$0.44 & 0.33$\pm$0.04 &    6.41 &  1.61 \\
 20203 &               & 274038922   &    8.85 & 7970  & 8.06 $\pm$0.38 & 0.45$\pm$0.05 &    7.20 &  1.76 &\cite{Holdsworth++2014}\\
 20232 &               & 159895674   &    6.88 & 8060  & 8.64 $\pm$0.36 & 0.44$\pm$0.05 &    6.86 &  1.64 \\
 24572 &               & 242944780   &    9.45 & 7410  & 7.25 $\pm$0.36 & 0.35$\pm$0.05 &    7.20 &  1.58 &\cite{Holdsworth++2014}\\
 24975 &               &  44645679   &    7.24 & 7790  & 9.20 $\pm$0.39 & 0.35$\pm$0.04 &    6.23 &  1.58 \\
 25248 &               & 459942890   &    8.60 & ---   & \emdash        & \emdash       &    7.16 &  1.55 \\
 25369 &               &   9147509   &    9.68 & ---   & \emdash        & \emdash       &    6.15 &  1.65 \\
 25674 &               &  34197596   &    8.69 & 8260  & 10.20$\pm$0.50 & 0.42$\pm$0.05 &    6.75 &  1.79 &\cite{Holdsworth++2014}\\
 28548 &               &  71134596   &    9.22 & 8510  & 10.82$\pm$0.55 & 0.45$\pm$0.06 &    7.56 &  1.67 &\cite{Holdsworth++2014,Gray++2017}\\
 29783 &               & 269792989   &    7.87 & ---   & \emdash        & \emdash       &    6.74 &  1.79 &\\
 30422 &       EX~Eri  &    589826   &    6.18 & 7940  & 8.42 $\pm$0.35 & 0.42$\pm$0.05 &    6.52 &  1.86 &\cite{Rodriguez2000,Murphy++2015}\\
 31322 &               & 246902545   &    9.28 & 8260  & 13.19$\pm$0.67 & 0.32$\pm$0.04 &    6.10 &  1.69 &\cite{Holdsworth++2014}\\
 31640 &               & 259675399   &    8.06 & 7690  & 8.25 $\pm$0.35 & 0.37$\pm$0.05 &    6.41 &  1.75 \\
 31901 &               & 316920092   &    9.07 & 7770  & 7.74 $\pm$0.39 & 0.41$\pm$0.05 &    6.97 &  1.56 &\cite{Meingast++2019}\\
 32433 &               & 348792358   &    9.22 & 7700  & 7.32 $\pm$0.35 & 0.42$\pm$0.05 &    6.95 &  1.54 \\
 34282 &    V1366~Ori  &  24344701   &    9.92 & ---   & \emdash        & \emdash       &    7.40 &  1.71 &\cite{Amado++2004,Casey++2013-HD34282}\\
 37286 &      HR~1915  &  31475829   &    6.26 & 8080  & 8.18 $\pm$0.34 & 0.47$\pm$0.06 &    7.30 &  ---  &\cite{Zuckerman++2011}\\ 
 38597 &               & 100531058   &    8.65 & 8430  & 10.38$\pm$0.47 & 0.44$\pm$0.05 &    6.90 &  1.68 &\cite{Holdsworth++2014}\\
 38629 &               &  32763133   &    8.92 & 8170  & 11.27$\pm$0.53 & 0.35$\pm$0.04 &    7.35 &  1.77 &\cite{Holdsworth++2014}\\
 39060 &  $\beta$~Pic  & 270577175   &    3.85 & 8080  & 8.49 $\pm$0.39 & 0.45$\pm$0.05 &    6.95 &  ---  &\cite{Mamajek+Bell2014,Mekarnia2017,Zwintz++2019}\\
 40317 &               & 282265535   &    8.45 & 8700  & 10.58$\pm$0.55 & 0.51$\pm$0.06 &    6.95 &  1.63 \\
 42005 &               & 408906554   &    9.54 & 8030  & 8.75 $\pm$0.42 & 0.42$\pm$0.05 &    7.16 &  1.57 &\cite{Holdsworth++2014}\\
 42608 &               &  37498433   &    9.85 & 8170  & 10.05$\pm$0.49 & 0.40$\pm$0.05 &    7.10 &  1.57 &\cite{Holdsworth++2014}\\
 42915 &               & 124429243   &    9.04 & 8520  & 12.82$\pm$0.68 & 0.38$\pm$0.05 &    6.40 &  ---  &\cite{Holdsworth++2014,Torres2006,Murphy+Lawson2015}\\ 
 44726 &               & 150272131   &   10.38 & 7890  & 7.87 $\pm$0.38 & 0.44$\pm$0.05 &    7.25 &  1.70 &\cite{Holdsworth++2014}\\
 44930 &               &  34737955   &    9.42 & 7320  & 7.17 $\pm$0.40 & 0.33$\pm$0.05 &    6.03 &  1.67 &\cite{Gray++2017}\\
 44958 &     V435~Car  & 255548143   &    6.74 & 7660  & 7.82 $\pm$0.32 & 0.38$\pm$0.05 &    6.90 &  1.58 &\cite{Rodriguez2000}\\
 45424 &               & 117766204   &    7.18 & 8060  & 10.39$\pm$0.44 & 0.36$\pm$0.04 &    6.54 &  1.86 \\
 46722 &               & 172193026   &    9.29 & 7810  & 8.28 $\pm$0.40 & 0.40$\pm$0.05 &    6.45 &  1.71 &\cite{Gray++2017}\\
 48985 &               & 148228220   &    9.04 & 7710  & 11.60$\pm$0.54 & 0.25$\pm$0.03 &    7.25 &  1.56 \\
 50153 &               &  78492107   &    7.03 & 7820  & 9.15 $\pm$0.39 & 0.36$\pm$0.05 &    6.85 &  1.60 \\
 54711 &               & 284348793   &    9.01 & 8200  & 9.22 $\pm$0.45 & 0.44$\pm$0.06 &    7.08 &  1.65 \\
 55863 &               & 294157254   &    9.06 & 7650  & 7.80 $\pm$0.38 & 0.38$\pm$0.05 &    6.92 &  1.57 \\
 59104 &               & 278179191   &    8.50 & 7360  & 6.15 $\pm$0.26 & 0.41$\pm$0.05 &    6.96 &  1.67 \\
 59594 &     V349~Pup  & 112484997   &    7.32 & 7800  & 8.06 $\pm$0.34 & 0.40$\pm$0.05 &    6.65 &  1.61 &\cite{Rodriguez2000}\\
 67688 &               & 306773428   &    7.66 & ---   & \emdash        & \emdash       &    7.04 &  1.59 \\
 70510 &               & 388351327   &    6.75 & ---   & \emdash        & \emdash       &    7.16 &  1.68 \\
 75040 &               & 176400189   &    9.05 & ---   & \emdash        & \emdash       &    6.64 &  1.58 \\
 78198 &               & 332516661   &    9.50 & 7340  & 7.79 $\pm$0.42 & 0.31$\pm$0.04 &    5.90 &  1.65 \\
 89263 &      HR~4043  & 463556278   &    6.22 & ---   & \emdash        & \emdash       &    7.02 &  1.71 &\\
 99506 &               & 287347434   &    8.36 & 7970  & 7.58 $\pm$0.37 & 0.48$\pm$0.05 &    7.05 &  1.59 &\cite{Holdsworth++2014}\\
220811 &               & 327996759   &    6.91 & ---   & \emdash        & \emdash       &    6.04 &  1.82 \\
222496 &               & 316806320   &    9.48 & ---   & \emdash        & \emdash       &    5.63 &  1.71 \\
223011 &               & 224284988   &    6.32 & 7830  & 10.49$\pm$0.44 & 0.31$\pm$0.04 &    5.93 &  1.68 \\
290750 &               &  11199304   &    9.77 & 9170  & 19.14$\pm$1.13 & 0.35$\pm$0.05 &    5.96 &  ---  \\ 
290799 &    V1790~Ori  &  11361473   &   10.67 & 8780  & 13.21$\pm$0.98 & 0.42$\pm$0.06 &    7.55 &  1.65 &\cite{Paunzen++2015,Murphy++2015}\\
       &   SAO~150524  & 143381070   &    9.46 & 8030  & 7.88 $\pm$0.39 & 0.48$\pm$0.06 &    7.63 &  ---  \\
       &   SAO~249859  & 349645354   &    9.79 & 7070  & 7.99 $\pm$0.38 & 0.23$\pm$0.04 &    5.34 &  1.69 \\
       &TYC~85-867-1   & 431695696   &    9.63 & 7961  & 8.85 $\pm$0.57 & 0.40$\pm$0.05 &    7.26 &  1.61 \\
       &TYC~5945-497-1 & 124381332   &    9.69 & 8270  & 10.02$\pm$0.53 & 0.42$\pm$0.05 &    7.45 &  1.61 &\cite{Holdsworth++2014}\\
       &TYC~8533-329-1 & 260161111   &   10.70 & 8370  & 9.33 $\pm$0.51 & 0.48$\pm$0.06 &    7.27 &  ---  &\cite{Holdsworth++2014}\\
       &TYC~8564-537-1 & 340358522   &   10.59 & 7490  & 7.30 $\pm$0.36 & 0.37$\pm$0.05 &    7.15 &  1.54 \\
187547 &               & KIC~7548479 &    8.40 & 7470  & 6.74 $\pm$0.29 & 0.40$\pm$0.05 &    7.00 &  1.61 &\cite{Antoci++2011,Antoci++2014}\\
       &TYC~3132-1272-1& KIC~8415752 &   10.67 & 7780  & 9.83 $\pm$0.52 & 0.32$\pm$0.05 &    6.20 &  1.63 \\
       &               & KIC~9450940 &   12.68 & 7920  & 8.59 $\pm$0.58 & 0.41$\pm$0.06 &    6.15 &  1.67 \\
\noalign{\smallskip}
\hline
\noalign{\smallskip}
\end{supertabular}

\fi
\end{table}

\clearpage

\begin{table}
\small
\caption{\sffamily\textbf{Projected rotational velocities from high-resolution
  spectroscopy}\label{tab:vsini}}
\smallskip
\ifarxiv
\\[1ex]
{\sffamily\small
Values of projected radial velocity, \sfvsini, in parentheses indicate a close binary (see Methods), meaning the measurement may not be reliable. Details of sources (rightmost column) are given in Methods and refs.\cite{Royer++2007,Mora++2001,Royer++2002,Schroeder++2009,Antoci++2011}.}
\\[1ex]

\tablehead{\multicolumn{1}{c}{HD} & \multicolumn{1}{c}{Name} & \multicolumn{1}{c}{TIC} & \multicolumn{1}{c}{\sfvsini} & source \\
 & & & \multicolumn{1}{c}{(\sfkms)} & \\ 
\hline
\noalign{\smallskip}
}
\sffamily\footnotesize
\begin{supertabular}{rrrcl}
    2280 &                & 281499618       & $\sf~~26.4 \pm 1.3$ & AAT+Veloce \\
    3622 &                &  43363194       & $\sf~~50\pm6$ & LCO+NRES \\
   10779 &                & 229139161       & $\sf~~91 \pm 5$ & AAT+Veloce \\
   10961 &                & 231014033\rlap{}& $\sf~~(33 \pm 3)$ & AAT+Veloce \\
   17693 &                & 122686610       & $\sf~~14 \pm 1$ & AAT+Veloce \\
   20203 &                & 274038922       & $\sf~~40\pm25$ & SALT+RSS \\
   20232 &                & 159895674       & $\sf~~37 \pm 3$ & AAT+Veloce \\
   24975 &                &  44645679       & $\sf~~88 \pm 4$ & AAT+Veloce \\
   25674 &                &  34197596       & $\sf160\pm35$ & SALT+RSS \\
   28548 &                &  71134596       & $\sf200\pm50$ & WHT+ISIS\\
   30422 &       EX~Eri   &    589826       & $\sf128$ & literature\cite{Royer++2007} \\
   31322 &                & 246902545       & $\sf200\pm50$ & SALT+RSS \\
   31640 &                & 259675399       & $\sf136\pm 4$ & AAT+Veloce \\
   31901 &                & 316920092       & $\sf~~33\pm 4$ & LCO+NRES     \\
   34282 &    V1366~Ori   &  24344701       & $\sf129\pm11$ & literature\cite{Mora++2001} \\
   37286 &      HR~1915   &  31475829       & $\sf~~70$ & literature\cite{Royer++2002} \\
   38597 &                & 100531058       & $\sf150\pm40$ & SALT+RSS\\
   38629 &                &  32763133       & $\sf160\pm40$ & SALT+RSS \\
   39060 &  $\beta$~Pic   & 270577175       & $\sf122$ & literature\cite{Schroeder++2009} \\
   42005 &                & 408906554       & $\sf130\pm30$ & SALT+RSS \\
   42608 &                &  37498433       & $\sf~~41\pm1$ & Keck+HIRES \\
   42915 &                & 124429243       & $\sf118\pm5$  & Keck+HIRES \\
   44726 &                & 150272131       & $\sf130\pm40$ & SALT+RSS \\
   44958 &     V435~Car   & 255548143       & $\sf114\pm11$ & LCO+NRES \\
   48985 &                & 148228220       & $\sf~~40 \pm 4$ & AAT+Veloce \\ 
   54711 &                & 284348793       & $\sf~~50 \pm 2$ & AAT+Veloce \\
   55863 &                & 294157254       & $\sf~~99 \pm 5$ & AAT+Veloce \\
   70510 &                & 388351327\rlap{}& $\sf~~(94\pm10)$ & LCO+NRES \\
   75040 &                & 176400189\rlap{}& $\sf~~(24\pm3)$ & Keck+HIRES \\
   78198 &                & 332516661       & $\sf~~45\pm1$  & Keck+HIRES \\
   89263 &      HR~4043   & 463556278\rlap{}& $(\sf100\pm7)$ & Keck+HIRES \\
   99506 &                & 287347434       & $\sf~~26\pm2$ & Keck+HIRES \\
  220811 &                & 327996759\rlap{}& $(\sf261\pm40)$ & Keck+HIRES \& LCO+NRES \\
  223011 &                & 224284988       & $\sf~~43\pm2$ & LCO+NRES \\
         &TYC~5945-497-1  & 124381332       & $\sf178\pm37$ & Keck+HIRES \\
         &TYC~8533-329-1  & 260161111       & $\sf~~100\pm30$ & SALT+RSS \\
  187547 &                & KIC~7548479     & $\sf~~10\pm2$ & literature\cite{Antoci++2011} \\
         & TYC~3132-1272-1& KIC~8415752     & $\sf~~18\pm1$ & Keck+HIRES \\
         &                & KIC~9450940     & $\sf~~10\pm1$ & Keck+HIRES \\
\noalign{\smallskip}
\hline
\noalign{\smallskip}
\end{supertabular}

\fi
\end{table}

\end{document}